\documentstyle[twoside,fleqn,espcrc2]{article}

\newcommand{\be}{\begin{equation}}
\newcommand{\ee}{\end{equation}}

\newcommand{\ba}{\begin{eqnarray}}
\newcommand{\ea}{\end{eqnarray}}

\newcommand{\Tr}{{\rm Tr}}
\newcommand{\Ucr}{U^{\dagger}}
\newcommand{\Xcr}{X^{\dagger}}

\newcommand{\gru}{U(L)\otimes U(L)}
\newcommand{\La}{{\cal L}}
\newcommand{\sgru}{SU(L)\otimes SU(L)}
\newcommand{\I}{{\bf I}}
\newcommand{\csti}{F^{2}_{\pi}+LF^{2}_{X}}

\newcommand{\unme}{\frac{1}{2}}
\newcommand{\qu}{{\bf Q}}
\newcommand{\eps}{\varepsilon^{\mu\nu\rho\sigma}}

\newcommand{\cstit}{F^{2}_{\pi}+3F^{2}_{X}}

\title{The problem of the $U(1)$ axial symmetry\\
       and the chiral transition in QCD}

\author{E. Meggiolaro\address{Dipartimento di Fisica, Universit\`a di Pisa,
        and INFN, Sezione di Pisa, I--56127 Pisa, Italy\\
	E-mail: enrico.meggiolaro@df.unipi.it
}}
\begin{document}

\begin{abstract}

We discuss the role of the $U(1)$ axial symmetry for the phase structure of
QCD at finite temperature. In particular, supported by recent lattice results,
we analyse a scenario in which a $U(1)$--breaking condensate survives across
the chiral transition.
This scenario can be consistently reproduced using an effective Lagrangian
model. The effects of the $U(1)$ chiral condensate on the slope of the
topological susceptibility in the full theory with quarks are studied.
Further information on the new $U(1)$ chiral order parameter is derived
from the study (at zero temperature) of the radiative decays of the ``light''
pseudoscalar mesons in two photons.

\end{abstract}

\maketitle

\section{Introduction}

\noindent
It is well known that at zero temperature the $SU(L) \otimes SU(L)$ chiral
symmetry, in a QCD with $L$ massless quarks, is broken spontaneously by the
non--zero value of the so--called {\it chiral condensate},
$\langle \bar{q}q \rangle \equiv \sum_{i=1}^L \langle \bar{q}_i q_i \rangle$, 
and the $L^2-1$ $J^P=0^-$ mesons
are just the Goldstone bosons associated with this breaking.
At high temperatures the thermal energy breaks up the $q\bar{q}$ condensate,
leading to the restoration of chiral symmetry above a certain critical
temperature $T_{ch}$, defined as the temperature at which the
condensate $\langle \bar{q}q \rangle$ goes to zero.
Instead, the role of the $U(1)$ axial symmetry \cite{Weinberg75,tHooft76} for
the finite temperature phase structure of QCD has been so far not well studied
and it is still an open question of hadronic physics.

In the ``Witten--Veneziano mechanism'' \cite{Witten79a,Veneziano79}
for the resolution of the $U(1)$ problem, a fundamental role is played by
the so--called ``topological susceptibility'' in a QCD without
quarks, i.e., in a pure Yang--Mills (YM) theory, in the large--$N_c$ limit
($N_c$ being the number of colours):
\be
A = \displaystyle\lim_{k \to 0}
\displaystyle\lim_{N_c \to \infty}
\left\{ -i \displaystyle\int d^4 x e^{ikx} \langle T Q(x) Q(0) \rangle
\right\},
\label{eqn1}
\ee
where $Q(x) = {g^2 \over 64\pi^2}\varepsilon^{\mu\nu\rho\sigma} F^a_{\mu\nu}
F^a_{\rho\sigma}$ is the so--called ``topological charge density''.
This quantity enters into the expression for the mass of the $\eta'$.
Therefore, in order to study the role of the $U(1)$ axial symmetry for the
full theory at non--zero temperatures, one should consider the YM topological
susceptibility $A(T)$ at a given temperature $T$, formally defined as in
Eq. (\ref{eqn1}), where now $\langle \ldots \rangle$ stands for the
expectation value in the full theory at the temperature $T$ \cite{EM1998}.

The problem of studying the behaviour of $A(T)$ as a function of the
temperature $T$ was first addressed, in lattice QCD,
in Refs. \cite{Teper86,EM1992a,EM1995b}.
Recent lattice results \cite{Alles-et-al.97} (obtained for the $SU(3)$
pure--gauge theory) show that the YM topological susceptibility $A(T)$
is approximately constant up to the critical temperature $T_{ch}$,
it has a sharp decrease above the transition, but it remains different
from zero up to $\sim 1.2~T_{ch}$. We recall that, in the Witten--Veneziano
mechanism \cite{Witten79a,Veneziano79}, a (no matter how small!) value
different from zero for $A$ is related to the breaking of the $U(1)$ axial
symmetry, since it implies the existence of a {\it would--be} Goldstone
particle with the same quantum numbers of the $\eta'$.

Another way to address the same question is to look at the behaviour at
non--zero temperatures of the susceptibilities related to the
propagators for the following meson channels \cite{Shuryak94}
(we consider for simplicity the case of $L=2$ light flavours):
the isoscalar ($I=0$) scalar channel $O_\sigma = \bar{q} q$;
the isovector ($I=1$) scalar channel
$\vec{O}_\delta = \bar{q} {\vec{\tau} \over 2} q$;
the isovector ($I=1$) pseudoscalar channel
$\vec{O}_\pi = i\bar{q} \gamma_5 {\vec{\tau} \over 2} q$;
the isoscalar ($I=0$) pseudoscalar channel $O_{\eta'} = i\bar{q} \gamma_5 q$.
Under $SU(2)$ chiral transformations, $\sigma$ is mixed with $\pi$
(and $\delta$ is mixed with $\eta'$).
On the contrary, under $U(1)$ chiral transformations, $\pi$ is mixed
with $\delta$ (and $\sigma$ is mixed with $\eta'$).
In practice, one can construct, for each meson channel $f$, the
corresponding chiral susceptibility
\be
\chi_f = \displaystyle\int d^4x~ \langle T O_f(x) O_f^\dagger(0) \rangle ,
\label{eqn2}
\ee
and then define two order parameters:\\
$\chi_{SU(2) \otimes SU(2)} \equiv \chi_\sigma - \chi_\pi$, and
$\chi_{U(1)} \equiv \chi_\delta - \chi_\pi$.
If an order parameter is non--zero in the chiral limit, then the
corresponding symmetry is broken.
Present lattice data for these quantities seem to indicate that the $U(1)$
order parameter survives across $T_{ch}$, up to $\sim 1.2~T_{ch}$,
where the $\delta$--$\pi$ splitting is small but still
different from zero \cite{Bernard-et-al.97,Karsch00,Vranas00}.
In terms of the left--handed and right--handed quark
fields ($q_{L,R} \equiv {1 \over 2} (1 \pm \gamma_5) q$, with $\gamma_5 \equiv
-i\gamma^0\gamma^1\gamma^2\gamma^3$)
one has the following expression for the difference between the correlators
for the $\delta^+$ and $\pi^+$ channels:
\ba
\lefteqn{
\langle O_{\delta^+}(x) O_{\delta^+}^\dagger(0)
\rangle - \langle O_{\pi^+}(x) O_{\pi^+}^\dagger(0) \rangle =
} \nonumber \\
\lefteqn{
2 \left[ \langle \bar{u}_R d_L(x) \bar{d}_R u_L(0) \rangle
+ \langle \bar{u}_L d_R(x) \bar{d}_L u_R(0) \rangle \right] . }
\label{eqn3}
\ea
(The integral of this quantity is just equal to the $U(1)$ chiral
susceptibility $\chi_{U(1)} = \chi_\delta - \chi_\pi$.)
What happens below and above $T_{ch}$?
Below $T_{ch}$, in the chiral limit $\sup(m_i) \to 0$, the left--handed
and right--handed components of a given light quark flavour ({\it up} or
{\it down}, in our case with $L=2$) can be connected through the $q\bar{q}$
chiral condensate, giving rise to a non--zero contribution to the
quantity (\ref{eqn3}) (i.e., to the quantity $\chi_{U(1)}$).
But above $T_{ch}$ the $q\bar{q}$ chiral condensate is zero:
so, how can the quantity (\ref{eqn3}) (i.e., the quantity $\chi_{U(1)}$) be
different from zero also above $T_{ch}$, as indicated by present lattice data?
The only possibility in order to solve this puzzle seems to be that of
requiring the existence of a genuine four--fermion local condensate,
which is an order parameter for the $U(1)$ axial symmetry and which
remains different from zero also above $T_{ch}$.
This new condensate will be discussed in Section 2 and then we shall analyse
some interesting phenomenological consequences deriving from this hypothesis
\cite{EM2003}.

\section{The $U(1)$ chiral order parameter}

\noindent
Let us define the following temperatures:

$T_\chi$: the temperature at which the pure--gauge topological
susceptibility $A$ drops to zero. Present lattice results
indicate that $T_\chi \ge T_{ch}$ \cite{Alles-et-al.97}.

$T_{U(1)}$: the temperature at which the $U(1)$ axial symmetry
is (effectively) restored, meaning that, for $T>T_{U(1)}$, there are no
$U(1)$--breaking condensates.
The Witten--Veneziano mechanism implies that $T_{U(1)} \ge T_\chi$,
since, after all, the pure--YM topological susceptibility $A$ is a
$U(1)$--breaking condensate.
Moreover, if $\langle \bar{q} q \rangle \ne 0$ also the $U(1)$ axial symmetry
is broken, i.e., the chiral condensate is an order parameter also for
the $U(1)$ axial symmetry. Therefore we must have: $T_{U(1)} \ge T_{ch}$.
Present lattice results for the chiral susceptibilities indicate that
$T_{U(1)} > T_{ch}$ \cite{Bernard-et-al.97,Karsch00,Vranas00}.

Thus we need another quantity which could be an order parameter only for 
the $U(1)$ chiral symmetry \cite{EM1994a,EM1994b,EM1994c,EM1995a,EM2002a}.
The most simple quantity of this kind was found by 'tHooft
in Ref. \cite{tHooft76}.
For a theory with $L$ light quark flavours, it is a $2L$--fermion interaction
that has the chiral transformation properties of:
\be
{\cal L}_{eff} \sim \displaystyle{{\det_{st}}(\bar{q}_{sR}q_{tL})
+ {\det_{st}}(\bar{q}_{sL}q_{tR}) },
\label{eqn5}
\ee
where $s,t = 1, \ldots ,L$ are flavour indices, but the colour indices are 
arranged in a more general way (see Refs. \cite{EM1994c,EM1995a,EM2002a}).
It is easy to verify that ${\cal L}_{eff}$ is invariant under
$SU(L) \otimes SU(L) \otimes U(1)_V$, while it is not invariant under $U(1)_A$.
To obtain an order parameter for the $U(1)$ chiral symmetry, one can 
simply take the vacuum expectation value of ${\cal L}_{eff}$:
$C_{U(1)} = \langle {\cal L}_{eff} \rangle$.
The arbitrarity in the arrangement of the colour indices can be removed if we 
require that the new $U(1)$ chiral condensate is ``independent'' of the
usual chiral condensate $\langle \bar{q} q \rangle$, as explained in
Refs. \cite{EM1994c,EM1995a,EM2002a}. In other words, the condensate $C_{U(1)}$
is chosen to be a {\it genuine} $2L$--fermion condensate, with a zero
``disconnected part'', the latter being the contribution proportional
to $\langle \bar{q} q \rangle^L$,
corresponding to retaining the vacuum intermediate state in all the channels
and neglecting the contributions of all the other states.
As a remark, we observe that the condensate $C_{U(1)}$ so defined
turns out to be of order ${\cal O}(g^{2L - 2} N_c^L) = {\cal O}(N_c)$
in the large--$N_c$ expansion, exactly as the chiral condensate
$\langle \bar{q} q \rangle$.

The existence of a new $U(1)$ chiral order parameter has of course interesting
physical consequences, which can be revealed by analysing some relevant QCD
Ward Identities (WI's) (see Refs. \cite{EM1994b,EM2002a}).
In the case of the $SU(L) \otimes SU(L)$ chiral symmetry, one
immediately derives the following WI:
\be
\int d^4 x~ \langle T\partial^\mu A^a_\mu (x) 
i\bar{q} \gamma_5 T^b q(0) \rangle
= i \delta_{ab} {1 \over L} \langle \bar{q} q \rangle ,
\label{eqn6}
\ee
where $A^a_\mu = \bar{q}\gamma_\mu \gamma_5 T^a q$ are the $SU(L)$ axial
currents. If $\langle \bar{q} q \rangle \ne 0$ (in the chiral limit
$\sup(m_i) \to 0$), the anomaly--free WI (\ref{eqn6}) implies the existence
of $L^2-1$ non--singlet Goldstone bosons, interpolated by the hermitian fields
$O_b = i \bar{q} \gamma_5 T^b q$.
Similarly, in the case of the $U(1)$ axial symmetry, one finds that:
\be
\int d^4x~ \langle T\partial^\mu J_{5, \mu}(x) i\bar{q} \gamma_5 q(0)
\rangle = 2i \langle \bar{q} q \rangle ,
\label{eqn7}
\ee
where $J_{5, \mu}= {\bar{q} \gamma_\mu \gamma_5 q}$ is the $U(1)$ axial
current. But this is not the whole story! One also derives the following WI:
\be
\int d^4x~ \langle T\partial^\mu J_{5, \mu}(x) O_P(0) \rangle = 
2Li \langle {\cal L}_{eff}(0) \rangle ,
\label{EQN10.5}
\ee
where ${\cal L}_{eff}$ is the $2L$--fermion operator defined by
Eq. (\ref{eqn5}), while the hermitian field $O_P$ is defined as:
$O_P \sim i[ {\det} (\bar{q}_{sR}q_{tL}) - {\det} (\bar{q}_{sL}q_{tR}) ]$.
If the $U(1)$--breaking condensate survives across the chiral transition at
$T_{ch}$, i.e., $C_{U(1)} = \langle {\cal L}_{eff}(0) \rangle \ne 0$ for
$T > T_{ch}$ (while $\langle \bar{q} q \rangle = 0$ for $T > T_{ch}$), then
this WI implies the existence of a ({\it would--be}) Goldstone boson (in the
large--$N_c$ limit) coming from this breaking and interpolated by the hermitian
field $O_P$. Therefore, the $U(1)_A$ ({\it would--be}) Goldstone boson (i.e.,
the $\eta'$) is an ``exotic'' $2L$--fermion state for $T > T_{ch}$.

\section{The new chiral effective Lagrangian}

\noindent
It is well known that the low--energy dynamics of the pseudoscalar mesons, 
including the effects due to the anomaly and the $q\bar{q}$ chiral condensate,
can be described, in the large--$N_c$ limit, and expanding to the first order
in the light quark masses, by an effective Lagrangian
\cite{DiVecchia-Veneziano80,Witten80,Rosenzweig-et-al.80,Nath-Arnowitt81,Ohta80}
written in terms of the mesonic field $U_{ij} \sim \bar{q}_{jR} q_{iL}$
(up to a multiplicative constant) and the topological charge density $Q$.
We make the assumption that there is a $U(1)$--breaking condensate which
stays different from zero across $T_{ch}$, up to $T_{U(1)} > T_{ch}$:
the form of this condensate has been discussed in the previous section.
We must now define a field variable $X$, associated with this new condensate,
to be inserted in the chiral Lagrangian.
The operators $i \bar{q} \gamma_5 q$ and $\bar{q} q$ entering in the WI
(\ref{eqn7}) are essentially equal to (up to a multiplicative constant)
$i(\Tr U - \Tr U^\dagger)$ and $\Tr U + \Tr U^\dagger$ respectively.
Similarly, the form of the new field $X$, in terms of the fundamental quark
fields, can be derived from the WI (\ref{EQN10.5}), identifying the operators
$O_P$ and ${\cal L}_{eff}$ with (up to a multiplicative constant)
$i(X - X^\dagger)$ and $X + X^\dagger$ respectively: this gives
$X \sim {\det} \left( \bar{q}_{sR} q_{tL} \right)$
(up to a multiplicative constant).
It was shown in Refs. \cite{EM1994a,EM1994b,EM1994c,EM2002a}
that the most simple effective Lagrangian, constructed
with the fields $U$, $X$ and $Q$, is:
\ba
\lefteqn{
{\cal L}(U,U^\dagger ,X,X^\dagger ,Q) } \nonumber \\
& & = {1 \over 2}\Tr(\partial_\mu U\partial^\mu U^\dagger )
+ {1 \over 2}\partial_\mu X\partial^\mu X^\dagger \nonumber \\
& & -V(U,U^\dagger ,X,X^\dagger)
+ {1 \over 2}iQ\omega_1 \Tr(\ln U - \ln U^\dagger) \nonumber \\
& & + {1 \over 2}iQ(1-\omega_1)(\ln X-\ln X^\dagger) + {1 \over 2A}Q^2,
\label{eqn9}
\ea
where the potential term $V(U,U^{\dagger},X,X^{\dagger})$ has the form:
\ba
\lefteqn{
V(U,U^\dagger ,X,X^\dagger ) } \nonumber \\
& & = {\lambda_{\pi}^2 \over 4} \Tr[(U^\dagger U
-\rho_\pi {\bf I})^2] +
{\lambda_X^2 \over 4} (X^\dagger X-\rho_X )^2 \nonumber \\
& & -{B_m \over 2\sqrt{2}}\Tr(MU+M^\dagger U^\dagger) \nonumber \\
& & -{c_1 \over 2\sqrt{2}}[\det(U)X^\dagger + \det(U^\dagger )X].
\label{eqn10}
\ea
$M={\rm diag}(m_1,\ldots,m_L)$ is the quark mass matrix.
All the parameters appearing in the Lagrangian must be considered as 
functions of the physical temperature $T$. In particular, the parameters 
$\rho_{\pi}$ and $\rho_X$ determine the expectation values $\langle U \rangle$
and $\langle X \rangle$ and so they are responsible for the behaviour of the
theory respectively across the $SU(L) \otimes SU(L)$ and the $U(1)$ chiral
phase transitions, as follows:
\ba
\lefteqn{
\rho_\pi|_{T<T_{ch}} \equiv {1 \over 2} F_\pi^2 > 0, ~~~
\rho_\pi|_{T>T_{ch}} < 0; } \nonumber \\
\lefteqn{
\rho_X|_{T<T_{U(1)}} \equiv {1 \over 2} F_X^2 > 0, ~~~
\rho_X|_{T>T_{U(1)}} < 0. }
\label{table}
\ea
The parameter $F_\pi$ is the well--known pion decay constant, while the
parameter $F_X$ is related to the new $U(1)$ axial condensate and will be
the object of our analysis.
According to what we have said in the Introduction and in Section 2,
we also assume that the topological susceptibility $A(T)$ of the pure--YM
theory drops to zero at a temperature $T_{\chi} \ge T_{ch}$
(but $T_{\chi} \le T_{U(1)}$).

One can study the mass spectrum of the theory for $T < T_{ch}$ and
$T_{ch} < T < T_{U(1)}$. First of all, let us see what happens for
$T<T_{ch}$, where both the $q\bar{q}$ chiral condensate and the $U(1)$ chiral
condensate are present. Integrating out the field variable $Q$ and taking only
the quadratic part of the Lagrangian, one finds that, in the chiral limit
$\sup(m_i) \to 0$, there are $L^2-1$ zero--mass states, which represent the
$L^2-1$ Goldstone bosons coming from the breaking of the $SU(L) \otimes SU(L)$
chiral symmetry down to $SU(L)_V$. Then there are two singlet eigenstates
with non--zero masses:
\ba
\lefteqn{
\eta' = {1 \over \sqrt{F_\pi^2 + LF_X^2}}(\sqrt{L}F_X S_X + F_\pi S_\pi), }
\nonumber \\
\lefteqn{
\eta_X = {1 \over \sqrt{F_\pi^2 + LF_X^2}}(-F_\pi S_X + \sqrt{L}F_X S_\pi), }
\label{eqn11}
\ea
where $S_\pi$ is the usual ``quark--antiquark'' $SU(L)$--singlet meson field
associated with $U$, while $S_X$ is the ``exotic'' $2L$--fermion meson field
associated with $X$ \cite{EM1994a,EM1994c,EM2002a}:
\ba
\lefteqn{
U = \frac{F_\pi}{\sqrt2}\exp\left[ {i\sqrt{2}\over F_\pi}
\left( \displaystyle\sum_{a=1}^{L^2-1}
\pi_{a}\tau_{a}+\frac{S_{\pi}}{\sqrt L}\I \right) \right], }
\nonumber \\
\lefteqn{
X = \frac{F_X}{\sqrt2}\exp\left({i\sqrt{2}\over F_X} S_X\right). }
\label{u,x}
\ea
The matrices $\tau_a$ ($a=1,\ldots,L^2-1$) are the generators of the
algebra of $SU(L)$ in the fundamental representation, with normalization:
$\Tr(\tau_a \tau_b) = \delta_{ab}$. The $\pi_a$ are the self--hermitian
fields describing the $L^2-1$ massless pions.\\
The field $\eta'$ has a ``light'' mass, in the sense of the
$N_c \to \infty$ limit, being
\be
m^2_{\eta'} = {2LA \over F_\pi^2 + LF_X^2} = {\cal O}({1 \over N_c}).
\label{eqn12}
\ee
On the contrary, the field $\eta_X$ has a sort of ``heavy hadronic'' mass of
order ${\cal O}(N_c^0)$ in the large--$N_c$ limit.
Both the $\eta'$ and the $\eta_X$ have the same quantum numbers (spin, 
parity and so on), but they have a different quark content: one is mostly
$S_\pi \sim i(\bar{q}_{L}q_{R}-\bar{q}_{R}q_{L})$, while the other is mostly
$S_X \sim i[ {\det}(\bar{q}_{sL}q_{tR}) - {\det}(\bar{q}_{sR}q_{tL}) ]$.
What happens when approaching the chiral transition temperature $T_{ch}$?
We know that $F_\pi(T) \to 0$ when $T \to T_{ch}$. From Eq. (\ref{eqn12})
we see that $m^2_{\eta'}(T_{ch}) = {2A \over F_X^2}$
and, from the first Eq. (\ref{eqn11}), $\eta'(T_{ch}) = S_X$.
We have continuity in the mass spectrum of the theory through the chiral 
phase transition at $T=T_{ch}$.
In fact, if we study the mass spectrum of the theory in the region of
temperatures $T_{ch} < T < T_{U(1)}$ (where the $SU(L) \otimes SU(L)$ chiral
symmetry is restored, while the $U(1)$ chiral condensate is still present),
one finds that there is a singlet meson field $S_X$ (associated with the 
field $X$ in the chiral Lagrangian) with a squared mass given by (in the
chiral limit): $m^2_{S_X} = {2A \over F_X^2}$.
This is nothing but the {\it would--be} Goldstone particle 
coming from the breaking of the $U(1)$ chiral symmetry, i.e., the $\eta'$,
which, for $T>T_{ch}$, is a sort of ``exotic'' matter field of the form
$S_X \sim i[ {\det}(\bar{q}_{sL}q_{tR}) - {\det}(\bar{q}_{sR}q_{tL}) ]$.
Its existence could be proved perhaps in the near future by
heavy--ion experiments.
 
\section{A relation between $\chi'$ and the new $U(1)$ chiral condensate}

\noindent
In this section and in the following one we want to describe some methods which
provide us with some information about the parameter $F_X$ \cite{EM2003}.
This quantity is a $U(1)$--breaking parameter: indeed, from Eq. (\ref{table}),
$\rho_X = {1 \over 2} F_X^2 > 0$ for $T<T_{U(1)}$, and therefore, from Eq.
(\ref{eqn10}), $\langle X \rangle = F_X/\sqrt{2} \ne 0$. Remembering that
$X \sim {\det} \left( \bar{q}_{sR} q_{tL} \right)$, up to a multiplicative
constant, we find that $F_X$ is proportional to the new $2L$--fermion
condensate $C_{U(1)} = \langle {\cal L}_{eff} \rangle$ introduced above.\\
In the same way, the pion decay constant $F_{\pi}$, which controls the breaking
of the $\sgru$ symmetry, is related to the $q\bar{q}$ chiral condensate 
by a simple and well--known proportionality relation (see Refs.
\cite{EM1994a,EM2002a} and references therein):
$\langle \bar{q}_i q_i \rangle_{T<T_{ch}} \simeq -{1 \over 2}B_m F_\pi$.
Considering, for simplicity, the case of $L$ light quarks with the same mass
$m$, one immediately derives from this equation the so--called
\emph{Gell-Mann--Oakes--Renner relation}:
$m_{NS}^{2}F_{\pi}^{2}\simeq-\frac{2m}{L}\langle\bar{q}q\rangle_{T<T_{ch}}$,
where, as usual,
$\langle \bar{q}q \rangle \equiv \sum_{i=1}^L \langle \bar{q}_i q_i \rangle$,
and, moreover, $m^2_{NS} = m B_m/F_\pi$ is the squared mass of the
non--singlet pseudoscalar mesons.\\
It is not possible to find, in a simple way, the analogous relation between
$F_X$ and the new condensate $C_{U(1)} = \langle {\cal L}_{eff} \rangle$.

Alternatively, the quantity $F_{X}$ can be extracted from the
two--point Green function of the topological charge--density operator 
$Q(x)$ in the {\it full} theory with $L$ light quarks:
\be
\chi(k)\equiv-i\int d^{4}x\;e^{ikx}\langle{TQ(x)Q(0)}\rangle.
\label{chik}
\ee
The calculation of $\chi(k)$ can be performed explicitly, using our effective
Lagrangian. The most interesting result is found when considering the
so--called ``slope'' of the topological susceptibility, defined as:
\ba
\label{chiprimo}
\lefteqn{
\chi'\equiv\frac{1}{8}{\partial \over \partial k_\mu}
{\partial \over \partial k^\mu} \chi(k)\Bigg|_{k=0}=
\frac{d}{dk^{2}}\chi(k)\Bigg|_{k=0}
} \nonumber \\
& & = \frac{i}{8}\int d^{4}x \;x^{2} \langle TQ(x)Q(0)\rangle,
\ea
which, in the chiral limit of $L$ {\it massless} quarks, comes out to be,
for $T<T_{ch}$ \cite{EM2003}:
\ba
\label{chi'sotto}
\chi'_{ch}=-\frac{1}{2L}(\csti)\equiv-\frac{1}{2L}F_{\eta'}^2,
\ea
where $F_{\eta'}\equiv\sqrt{\csti}$ is the decay constant of the $\eta'$ (at 
the leading order in the $1/N_c$ expansion), modified by the presence of the 
new $U(1)$ chiral order parameter \cite{EM1994c,EM2002a}.
In fact, remembering how the fields $U$ and $X$ transform under a $U(1)$ chiral
transformation, one can determine the $U(1)$ axial current, starting from our
effective Lagrangian \cite{EM1994c,EM2002a}:
\ba
\lefteqn{
J_{5, \mu} = i \left[ \Tr(U^\dagger \partial_\mu U - U \partial_\mu U^\dagger)
\right. } \nonumber \\
\lefteqn{
\left. + L(X^\dagger \partial_\mu X - X \partial_\mu X^\dagger) \right]
= -\sqrt{2L} F_{\eta'} \partial_\mu \eta' ,}
\label{EQN5.15}
\ea
where the field $\eta'$ is defined by the first Eq. (\ref{eqn11}) and
the relative coupling between $J_{5, \mu}$ and $\eta'$, i.e., the
$SU(L)$--singlet ($\eta'$) decay constant defined as
$\langle 0|J_{5, \mu}(0)|\eta'(p)\rangle = i\sqrt{2L}\,p_\mu\,F_{\eta'}$,
comes out to be:
\be
F_{\eta'} = \sqrt{\csti} .
\label{EQN5.18}
\ee
Summarizing, we have found that the value of $\chi'$, in the chiral limit
$\sup(m_i)\to0$, is shifted from the ``original'' value
$-\frac{1}{2L}F_{\pi}^2$ (derived in the absence of an extra $U(1)$ chiral
condensate: see Refs. \cite{spin-crisis,GRTV02}) to the value
$-\frac{1}{2L}F_{\eta'}^2=-\frac{1}{2L}(\csti)$, which also depends on the
quantity $F_X$, proportional to the extra $U(1)$ chiral condensate.

All the above refers to the theory at $T<T_{ch}$.
When approaching the chiral transition at $T=T_{ch}$, one expects
that $F_{\pi}$ vanishes, while $F_X$ remains different from zero and the
quantity $\chi'_{ch}$ tends to the value:
\be
\chi'_{ch}\mathop{\longrightarrow}_{T\to T_{ch}}-\unme F_X^2.
\label{chi'Tch}
\ee
The quantity $\chi(k)$ can also be evaluated in the region of temperatures
$T_{ch}<T<T_{U(1)}$, proceeding as for the case $T<T_{ch}$, obtaining the
result (already derived in Ref. \cite{EM1994a}):
\be
\chi(k)=A\frac{k^{2}}{k^{2}-\frac{2A}{F^2_X}},
\ee
in the chiral limit $\sup(m_i)\to0$.\\
Therefore, in the region of temperatures $T_{ch}<T<T_{U(1)}$, $\chi'_{ch}$
is given by \cite{EM2003}:
\be
\label{chi'sopra}
\chi'_{ch}=\frac{d}{dk^{2}}\chi(k)\Bigg|_{k=0}=
-\frac{1}{2}\,F_X^{2},
\ee
consistently with the results (\ref{chi'sotto}) and (\ref{chi'Tch}) found
above: i.e., $\chi'_{ch}$ varies with continuity across $T_{ch}$.
This means that $\chi'_{ch}$ acts as a sort of order parameter for the $U(1)$
axial symmetry above $T_{ch}$: if $\chi'_{ch}$ is different from zero above
$T_{ch}$, this means that the $U(1)$--breaking parameter $F_X$ is different
from zero.

\section{Radiative decays of the pseudoscalar mesons}

\noindent
Further information on the quantity $F_X$ (i.e., on the new $U(1)$ chiral
condensate, to which it is related) can be derived from the study of the
radiative decays of the ``light'' pseudoscalar mesons in two photons,
$\pi^0,\eta,\eta',\eta_X\to\gamma\gamma$, in the realistic case of $L=3$
light quarks (with non--zero masses $m_u$, $m_d$ and $m_s$) and in the simple
case of zero temperature ($T=0$) \cite{EM2003}.\\
To this purpose, we have to introduce the electromagnetic interaction in our 
effective model. Under {\it local} $U(1)$ electromagnetic transformations:
\be
q \to q' = e^{i\theta e \qu}q, ~~~
A_\mu \to A'_\mu = A_\mu - \partial_\mu \theta,
\ee
the fields $U$ and $X$ transform as follows:
\be
U \to U' = e^{i\theta e \qu} U e^{-i\theta e \qu}, ~~~
X \to X' = X.
\ee
Therefore, we have to replace the derivative of the fields $\partial_\mu U$
and $\partial_\mu X$ with the corresponding {\it covariant} derivatives:
\be
D_{\mu}U=\partial_{\mu}U+ie A_{\mu}[\qu,U], ~~~ D_{\mu}X = \partial_\mu X.
\ee
Here $\qu$ is the quark charge matrix (in units of $e$, the absolute value of
the electron charge):
\be
\label{caricael}
\qu=
\left( \begin{array}{ccc}
\frac{2}{3} & \\
& -\frac{1}{3} \\
& & -\frac{1}{3} \\
\end{array}\right).
\ee
In addition, we have to reproduce the effects of the electromagnetic anomaly,
whose contribution to the four--divergence of the $U(1)$ axial current
($J_{5,\mu}=\bar{q}\gamma_{\mu}\gamma_{5}q$) and of the $SU(3)$ axial currents
($A^{a}_{\mu}=\bar{q}\gamma_{\mu}\gamma_{5}\frac{\tau_{a}}{\sqrt{2}}q$)
is given by:
\ba
\lefteqn{
(\partial^{\mu}J_{5,\mu})^{e.m.}_{anomaly} = 2\Tr(\qu^{2}) G, }
\nonumber \\
\lefteqn{
(\partial^{\mu}A^{a}_{\mu})^{e.m.}_{anomaly} =
2\Tr\left( \qu^{2}\frac{\tau_{a}}{\sqrt{2}}\right) G, }
\ea
where $G\equiv\frac{e^{2}N_{c}}{32\pi^{2}}\eps F_{\mu\nu}F_{\rho\sigma}$ 
($F_{\mu\nu}$ being the electromagnetic field--strength tensor), thus 
breaking the corresponding chiral symmetries. We observe that
$\Tr(\qu^{2}\tau_{a}) \ne 0$ only for $a=3$ or $a=8$.\\
We must look for an interaction term ${\cal L}_I$ (constructed with the chiral
Lagrangian fields and the electromagnetic operator $G$) which, under a $U(1)$
axial transformation $q \to q' = e^{-i\alpha\gamma_5}q$, transforms as:
\be
U(1)_A:~~{\cal L}_I \to {\cal L}_I + 2\alpha \Tr(\qu^2)G,
\label{prop-u1}
\ee
while, under $SU(3)$ axial transformations of the type $q \to q' = e^{-i\beta
\gamma_5 \tau_a/\sqrt{2}}q$ (with $a = 3,8$), transforms as:
\be
SU(3)_A:~~{\cal L}_I \to {\cal L}_I + 2\beta \Tr\left( \qu^2
{\tau_a \over \sqrt{2}} \right) G.
\label{prop-su3}
\ee
By virtue of the transformation properties of the fields $U$ and $X$ under a
$\gru$ chiral transformation \cite{EM1994a,EM2002a}, one can see that the most
simple term describing the electromagnetic anomaly interaction term is the
following one:
\be
\label{li}
\La_{I}=\unme iG\Tr[\qu^{2}(\ln U-\ln\Ucr)],
\ee
which is exactly the one originally proposed in Ref.
\cite{DiVecchia-Veneziano-et-al.81}.
Therefore, we have to consider the following effective chiral Lagrangian, 
which includes the electromagnetic interaction terms described
above \cite{EM2003}:
\ba
\label{lem}
\lefteqn{\La(U,\Ucr,X,\Xcr,Q,A^{\mu}) } \nonumber \\
& & = \frac{1}{2}\Tr(D_{\mu}UD^{\mu}\Ucr)+
\frac{1}{2}\partial_{\mu}X\partial^{\mu}\Xcr \nonumber\\
& & -V(U,\Ucr,X,\Xcr)+\frac{1}{2}iQ\;\omega_{1}\Tr(\ln U-\ln \Ucr) \nonumber\\
& & +\frac{1}{2}iQ(1-\omega_{1})(\ln X-\ln \Xcr)+\frac{1}{2A} Q^{2}
\nonumber\\
& & + \unme iG\Tr[\qu^{2}(\ln U-\ln\Ucr)]
-\frac{1}{4}F_{\mu\nu}F^{\mu\nu}.
\ea
The decay amplitude of the generic process ``$meson\to\gamma\gamma$''
is entirely due to the electromagnetic anomaly interaction term, which can
be written more explicitly as follows, in terms of the meson fields:
\be
\label{li1}
\La_{I}=-G\frac{1}{3F_{\pi}} \left( \pi_{3}+\frac{1}{\sqrt{3}}\pi_{8}+
\frac{2\sqrt{2}}{\sqrt{3}}S_{\pi} \right).
\ee
The fields  $\pi_{3},\pi_{8},S_{\pi},S_{X}$ mix together. 
However, neglecting the experimentally small mass difference between the 
quarks \emph{up} and \emph{down} (i.e., neglecting the experimentally small
violations of the $SU(2)$ isotopic spin), also $\pi_3$ becomes diagonal and can
be identified with the physical state $\pi^0$.
The fields $(\pi_8,S_\pi,S_X)$ can be written in terms of the eigenstates
$(\eta,\eta',\eta_{X})$ as follows:
\be
\pmatrix{ \pi_8 \cr S_\pi \cr S_X } = \mathbf{C}
\pmatrix{ \eta \cr \eta' \cr \eta_X },
\label{diag}
\ee
where $\mathbf{C}$ is the following $3\times3$ orthogonal matrix:
\ba
\label{cambio}
\mathbf{C}=
\left(\begin{array}{ccc}
\cos\tilde{\varphi} & -\sin\tilde{\varphi} & 0\\
&&\\
\sin\tilde{\varphi}\,\frac{F_{\pi}}{F_{\eta'}} &
\cos\tilde{\varphi}\,\frac{F_{\pi}}{F_{\eta'}} &
\frac{\sqrt{3}F_{X}}{F_{\eta'}}\\
&&\\
\sin\tilde{\varphi}\,\frac{\sqrt{3}F_{X}}{F_{\eta'}}  & 
\cos\tilde{\varphi}\,\frac{\sqrt{3}F_{X}}{F_{\eta'}} & 
-\frac{F_{\pi}}{F_{\eta'}}
\end{array} \right) .
\ea
Here $F_{\eta'}$ is defined according to Eq. (\ref{EQN5.18}), i.e.,
\be
F_{\eta'} \equiv \sqrt{\cstit},
\label{F-eta'}
\ee
and $\tilde{\varphi}$ is a mixing angle, which can be related to the masses 
of the quarks $m_u$, $m_d$, $m_s$, and therefore to the masses of the
octet mesons, by the following relation:
\be
\label{phitilde}
\tan\tilde{\varphi}
= \frac{F_{\pi}{F_{\eta'}}}{6\sqrt{2}A}(m_{\eta}^{2}-m_{\pi}^{2}),
\ee
where:
$m^{2}_{\pi}=2B\tilde{m}$ and $m_{\eta}^{2}=\frac{2}{3}B(\tilde{m}+2m_s)$,
with: $B \equiv \frac{B_{m}}{2F_{\pi}}$, $\tilde{m} \equiv \frac{m_u+m_d}{2}$.
With simple standard calculations, the following decay rates (in the real case
$N_c=3$) are derived \cite{EM2003}:
\ba
\lefteqn{
\Gamma(\pi^{0}\to\gamma\gamma) =
\frac{\alpha^{2}m_{\pi}^{3}}{64\pi^{3}F_{\pi}^{2}}, } \nonumber \\
\lefteqn{
\Gamma(\eta\to\gamma\gamma) =
\frac{\alpha^{2}m_{\eta}^{3}}{192\pi^{3}F_{\pi}^{2}}\Big(\cos\tilde{\varphi}+
2\sqrt2\sin\tilde{\varphi}\,\frac{F_{\pi}}{F_{\eta'}}\Big)^{2}, } \nonumber \\
\lefteqn{
\Gamma(\eta'\to\gamma\gamma) =
\frac{\alpha^{2}m_{\eta'}^{3}}{192\pi^{3}F_{\pi}^{2}}
\Big(2\sqrt2\cos\tilde{\varphi}\,
\frac{F_{\pi}}{F_{\eta'}}-\sin\tilde{\varphi}\Big)^{2}, } \nonumber \\
\lefteqn{
\Gamma(\eta_{X}\to\gamma\gamma) =
\frac{\alpha^{2}m_{\eta_{X}}^{3}}{8\pi^{3}F_{\pi}^{2}}
\Big(\frac{F_{X}}{F_{\eta'}}\Big)^{2}, }
\label{gamma}
\ea
where $\alpha=e^{2}/4\pi \simeq 1/137$ is the fine--structure constant.

If we put $F_X=0$ (i.e., if we neglect the new $U(1)$ chiral condensate),
the expressions written above reduce to the corresponding ones derived in
Ref. \cite{DiVecchia-Veneziano-et-al.81} using an effective Lagrangian which
includes only the usual $q\bar{q}$ chiral condensate (so there is no field
$\eta_X$!). The introduction of the new condensate (while leaving the
$\pi^0\to\gamma\gamma$ decay rate unaffected, as it must!) modifies 
the decay rates of $\eta$ and $\eta'$ (and, moreover, we also have to consider
the particle $\eta_X$). In particular, it modifies the $\eta'$ decay constant,
already in the chiral limit $\sup(m_i) \to 0$, according to Eq. (\ref{F-eta'}).

In conclusion, a study of the radiative decays $\eta\to\gamma\gamma$,
$\eta'\to\gamma\gamma$ and a comparison with the experimental data can 
provide us with further information about the parameter $F_X$ and the new 
exotic condensate.
Using the experimental values for the various quantities which appear in
the second and third Eq. (\ref{gamma}), i.e.,
\ba
\lefteqn{
F_\pi = 92.4(4) ~{\rm MeV}, }
\nonumber \\
\lefteqn{
m_\eta = 547.30(12) ~{\rm MeV}, }
\nonumber \\
\lefteqn{
m_{\eta'} = 957.78(14) ~{\rm MeV}, }
\nonumber \\
\lefteqn{
\Gamma(\eta\to\gamma\gamma) = 0.46(4) ~{\rm KeV}, }
\nonumber \\
\lefteqn{
\Gamma(\eta'\to\gamma\gamma) = 4.26(19) ~{\rm KeV}, }
\ea
we can extract the following values for the quantity $F_X$ and for the
mixing angle $\tilde\varphi$ \cite{EM2003}:
\be
F_X = 27(9) ~{\rm MeV},~~~ \tilde\varphi = 16(3)^0.
\ee
Moreover, the values of $F_X$ and $\tilde\varphi$ so
found are perfectly consistent with the relation (\ref{phitilde}) for the
mixing angle, if we use for the pure--YM topological susceptibility
the value $A=(180\pm5~\rm{MeV})^{4}$, obtained from lattice simulations.

\section{Conclusions}

\noindent
There are evidences from some lattice results that
a new $U(1)$--breaking condensate survives across the chiral transition
at $T_{ch}$, staying different from zero up to $T_{U(1)} > T_{ch}$.
This scenario can be consistently reproduced using an effective Lagrangian
model, which also includes the new $U(1)$ chiral condensate.
This scenario could perhaps be verified in the near future by heavy--ion
experiments, by analysing the pseudoscalar--meson spectrum in the singlet
sector.

We have determined the effects due to the presence of the new $U(1)$ chiral
order parameter on the slope of the topological susceptibility $\chi'_{ch}$,
in the {\it full} theory with $L$ {\it massless} quarks.
We have found that $\chi'_{ch}$ acts as an order parameter
for the $U(1)$ axial symmetry above $T_{ch}$ \cite{EM2003}.
This prediction of our model could be tested in the near--future Monte
Carlo simulations on the lattice: at present, lattice determinations of
$\chi'$ only exist for the pure--gauge theory at $T=0$, with gauge group
$SU(2)$ \cite{Briganti-et-al.91} and $SU(3)$ \cite{EM1992b}
(but see also Ref. \cite{GRTV02} for a discussion about possible ambiguities
in the definition of $\chi'_{ch}$ in a lattice regularized theory).

We have also investigated the effects of the new $U(1)$ chiral condensate on
the radiative decays (at $T=0$) of the pseudoscalar mesons $\eta$ and $\eta'$
in two photons. A first comparison of our results with the experimental data
has been performed: the results are encouraging, pointing towards a certain
evidence of a non--zero $U(1)$ axial condensate
(i.e., $F_X \ne 0$) \cite{EM2003}.

However, one should keep in mind that our results have been
derived from a very simplified model, obtained doing a first--order expansion
in $1/N_c$ and in the quark masses. We expect that such a model can furnish
only qualitative or, at most, ``semi--quantitative'' predictions.
When going beyond the leading order in $1/N_c$, it becomes
necessary to take into account questions of renormalization--group
behaviour of the various quantities and operators involved in our
theoretical analysis. This issue has been widely discussed in the
literature, both in relation to the analysis of $\chi'_{ch}$,
in the context of the proton--spin crisis problem \cite{spin-crisis},
and also in relation to the study of the $\eta,\eta'$ radiative decays
\cite{gamma-gamma}.
Further studies are therefore necessary in order to continue this analysis
from a more quantitative point of view.\\
Last, but not least, it would be also very interesting (for a comparison
with future heavy--ion experiments) to extend our present analysis of the
radiative decays to the non--zero--temperature case.
We expect that some progress will be done along this line in the near future.

\end{document}